\documentclass[showpacs,onecolumn,amssymb,prl,aps]{revtex4}

\usepackage{graphicx,epsfig}
\usepackage{dcolumn}
\usepackage{bm}
\begin{document}

\title{Exact analytical results for quantum walks on star graph}
\author{Xin-Ping Xu}
\affiliation{%
$^1$Institute of Particle Physics, HuaZhong Normal University, Wuhan
430079, China  \\
$^2$Institute of High Energy Physics, Chinese Academy of Science,
Beijing 100049, China
}%

\begin{abstract}
In this paper, we study coherent exciton transport of
continuous-time quantum walks on star graph. Exact analytical
results of the transition probabilities are obtained by means of the
Gram-Schmidt orthonormalization of the eigenstates. Our results show
that the coherent exciton transport displays perfect revivals and
strong localization on the initial node. When the initial excitation
starts at the central node, the transport on star graph is
equivalent to the transport on a complete graph of the same size.
\end{abstract}
 \maketitle
The problem of coherent and non-coherent transport modeled by random
walks has attracted much attention in many distinct fields, ranging
from polymer physics to biological physics, from solid state physics
to quantum computation~\cite{rn1,rn2,rn3,rn4}. Random walk is
related to the diffusion models and a fundamental topic in
discussions of Markov processes. Several properties of random walks,
including dispersal distributions, first-passage times and encounter
rates, have been extensively studied~\cite{rn5,rn6}. As a natural
extension to the quantum world of the ubiquitous classical random
walks, quantum walks (QWs) have also been introduced and widely
investigated in the literature~\cite{rn7}.

An important application of quantum walks is that QWs can be used to
design highly efficient quantum algorithms. For example, Grover¡¯s
algorithm can be combined with quantum walks in a quantum algorithm
for ¡°glued trees¡± which provides even an exponential speed up over
classical methods~\cite{rn8,rn9}. Besides their important
applications in quantum algorithms, quantum walks are also used to
model the coherent exciton transport in solid state
physics~\cite{rn10}. It is shown that the dramatic nonclassical
behavior of quantum walks can be attributed to quantum coherence,
which does not exist in the classical random walks.

There are two main types of quantum walks: continuous-time and
discrete-time quantum walks. The main difference between them is
that discrete-time walks require a ¡°coin¡±¡ª which is just any
unitary matrix¡ªplus an extra Hilbert space on which the coin acts,
while continuous-time walks do not need this extra Hilbert
space~\cite{rn11}. Apart from this difference, the two types of
quantum walks are analogous to continuous-time and discrete-time
random walks in the classical case~\cite{rn11}. Discrete-time
quantum walks evolve by the application of a unitary evolution
operator at discrete-time intervals, and continuous-time quantum
walks evolve under a time-independent Hamiltonian. Unlike the
classical case, the discrete-time and continuous-time quantum walks
cannot be simply related to each other by taking a limit as the time
step goes to zero~\cite{rn12}. Both the two types of quantum walks
have been defined and studied on discrete structures in the market.

Here, we focus on continuous-time quantum walks (CTQWs). Most of
previous studies related to CTQWs in the last decade concentrates on
regular structures such as lattices, and most of the common wisdom
concerning them relies on the results obtained in this particular
geometry. Exact analytical results for CTQWs are also found for some
particular regular structures, such as the cycle graph~\cite{rn13}
and Cayley tree~\cite{rn14}. Exact analytical results are difficult
to get due to the cumbersome analytical investigations of the
eigenvalues and eigenstates of the Hamiltonian.

In this paper, we consider CTQWs on star graphs, and get exact
analytical results for the first time. The star graph is one of the
most regular structures in the graph theory and represents the local
tree structure of the irregular and complex graphs. In mathematical
language, a star graph of size $N$ consists of one central node and
$N-1$ leaf nodes. All the leaf nodes connect to the central node,
and there is no connection between the leaf nodes. Therefore, the
central node has $N-1$ bonds and the leaf nodes have only one bond.
As we will shown, for such simple topology, we are able to derive
exact analytical results for the transition probabilities. These
analytical results exactly agree with the numerical results obtained
by diagonalizing the Hamiltonian $H$ using the software MATLAB.

The coherent exciton transport on a connected network is modeled by
the continuous-time quantum walks (CTQWs), which is obtained by
replacing the Hamiltonian of the system by the classical transfer
matrix, i.e., $H=-T$~\cite{rn15,rn16}. The transfer matrix $T$
relates to the Laplace matrix by $T=-A$. The Laplace matrix $A$ has
nondiagonal elements $A_{ij}$ equal to $-1$ if nodes $i$ and $j$ are
connected and $0$ otherwise. The diagonal elements $A_{ii}$ equal to
degree of node $i$, i.e., $A_{ii}=k_i$. The states $|j\rangle$
endowed with the node $j$ of the network form a complete,
ortho-normalised basis set, which span the whole accessible Hilbert
space. The time evolution of a state $|j\rangle$ starting at time
$t_0$ is given by $|j,t\rangle = U(t,t_0)|j\rangle$, where
$U(t,t_0)=exp[-iH(t-t_0)]$ is the quantum mechanical time evolution
operator. The transition amplitude $\alpha_{k,j}(t)$ from state
$|j\rangle$ at time $0$ to state $|k\rangle$ at time $t$ reads
$\alpha_{k,j}(t)=\langle k|U(t,0)|j\rangle$ and obeys
Schr\"{o}dinger¡¯s equation~\cite{rn17}. Then the classical and
quantum transition probabilities to go from the state $|j\rangle$ at
time $0$ to the state $|k\rangle$ at time $t$ are given by
$p_{k,j}(t)=\langle k|e^{-tA}|j\rangle$ and
$\pi_{k,j}(t)=|\alpha_{k,j}(t)|^2= |\langle
k|e^{-itH}|j\rangle|^2$~\cite{rn18}, respectively. Using $E_n$ and
$|q_n\rangle$ to represent the $n$th eigenvalue and ortho-normalized
eigenvector of $H$, the classical and quantum transition
probabilities between two nodes can be written as~\cite{rn17,rn18}
\begin{equation}\label{eq1}
p_{k,j}(t)=\sum_n e^{-tE_n}\langle k|q_n\rangle \langle
q_n|j\rangle,
\end{equation}
\begin{equation}\label{eq2}
\pi_{k,j}(t)=|\alpha_{k,j}(t)|^2=|\sum_n e^{-itE_n}\langle
k|q_n\rangle \langle q_n|j\rangle|^2.
\end{equation}

For finite networks, the classical transition probabilities approach
the equal-partition $1/N$. However, the quantum transport does not
lead to equal-partition. $\pi_{k,j}(t)$ do not decay ad infinitum
but at some time fluctuates about a constant value. This value is
determined by the long time average of
$\pi_{k,j}(t)$~\cite{rn17,rn18},
\begin{equation}\label{eq3}
\begin{array}{ll}
\chi_{k,j}&=\lim_{T\rightarrow \infty}\frac{1}{T}\int_0^T
\pi_{k,j}(t)dt\\
&=\sum_{n,l}\langle k|q_n\rangle \langle q_n|j\rangle \langle j|q_l\rangle \langle q_l|k\rangle \\
&\  \ \  \ \times\lim_{T\rightarrow \infty}\frac{1}{T}\int_0^T e^{-it(E_n-E_l)}dt\\
&=\sum_{n,l}\delta_{E_n,E_l}\langle k|q_n\rangle \langle
q_n|j\rangle \langle j|q_l\rangle \langle q_l|k\rangle.
\end{array}
\end{equation}
where $\delta_{E_n,E_l}$ takes value 1 if $E_n$ equals to $E_l$ and
0 otherwise. In order to calculate $p_{k,j}(t)$, $\pi_{k,j}(t)$ and
$\chi_{k,j}$, all the eigenvalues $E_n$ and eigenstates
$|q_n\rangle$ are required. For the star graph, in the following, we
will first analytically calculate the eigenvalues and eigenstates,
then give exact analytical results for the transition probabilities
according to the above Equations.

For a specific star graph of size $N$, we label the central node as
node $1$ while the leaf nodes are numbered as $2,3,...,N$. Because
of centrosymmetric structure of the star graph, there are only four
types of transition probabilities, namely, $\pi_{1,1}(t)$,
$\pi_{2,1}(t)\equiv \pi_{1,2}(t)$, $\pi_{2,2}(t)$ and
$\pi_{3,2}(t)\equiv \pi_{2,3}(t)$. Transition probabilities between
other nodes belong to these four types. Therefore, we only consider
the four kinds of probabilities listed above. The Hamiltonian of the
star graph can be written as,
\begin{equation}\label{eq4}
H=(N-1)|1\rangle \langle 1|+\sum_{i=2}^N(|i\rangle \langle
i|-|1\rangle \langle i|-|i\rangle \langle 1|).
\end{equation}
The eigenvalues of the Hamiltonian have three discrete values:
$E_1=E_2=...=E_{N-2}=1$, $E_{N-1}=0$ and $E_{N}=N$~\cite{add-ref}.
One set of eigenstates $\{|v_i\rangle\}$ ($i=1,2,...,N$)
corresponding to the eigenvalues is : $|v_i\rangle =|i+2\rangle
-|2\rangle$ ($i=1,2,...,N-2$), $|v_{N-1}\rangle=\sum_{i=1}^N
|i\rangle$ and $|v_{N}\rangle=\sum_{i=1}^N |i\rangle-N|1\rangle$.
However, this set of eigenstates are not orthogonal ($\langle
v_1|v_2\rangle \neq 0$, etc), we use the Gram-Schmidt
process~\cite{rn19} to orthogonalize this set of eigenstates. The
Gram-Schmidt algorithm is a method for orthogonalizing a set of
vectors in an inner product space~\cite{rn20}, the new orthogonal
vectors $\{|v_i'\rangle\} (i=1,2,...,N-2)$ are given by the
following formula,
\begin{equation}\label{eq5}
|v_i'\rangle=|v_i\rangle-\sum_{j=1}^{i-1}\frac{\langle
v_i|v_j'\rangle}{\langle v_j'|v_j'\rangle}|v_j'\rangle.
\end{equation}
where $|v_1'\rangle=|v_1\rangle$ is applied in the iterative
process~\cite{rn19,rn20}. According to the above equation, we
obtain,
\begin{equation}\label{eq6}
|v_i'\rangle=\left\{
\begin{array}{ll}
|i+2\rangle-\frac{1}{i}\sum_{j=2}^{i+1}|j\rangle,   &  i=1,2,...,N-2 \\
|v_i\rangle,  & i=N-1, N.
\end{array}
\right.
\end{equation}
The above new eigenstates $\{v_i'\rangle\}$ is not normalized (i.e.,
$\langle v_i'|v_i'\rangle\neq 1$, etc). After some algebraic
calculations, we get the orthonormal basis $\{|q_i\rangle\}$ as
follows,
\begin{equation}\label{eq7}
|q_i\rangle=\left\{
\begin{array}{lll}
\sqrt{\frac{i}{i+1}}|i+2\rangle-\sqrt{\frac{1}{i(i+1)}}\sum_{j=2}^{i+1}|j\rangle,   &  i=1,2,...,N-2 \\
\sqrt{1/N}\sum_{j=1}^N |j\rangle,  & i=N-1 \\
\frac{1}{\sqrt{N(N-1)}}\sum_{i=1}^N
|i\rangle-\sqrt{\frac{N}{N-1}}|1\rangle, & i=N.
\end{array}
\right.
\end{equation}

One can easily prove $\{|q_i\rangle\}$ is also a set of eigenstates
of the Hamiltonian (i.e., $H|q_i\rangle=E_i|q_i\rangle, \forall \
i$). $\{|q_i\rangle\}$ forms an orthonormal and complete basis,
satisfying $\langle q_i|q_j\rangle =\delta_{ij}$ and
$\sum_j|q_j\rangle\langle q_j|=1$. Therefore, we can use this set of
eigenstates $\{|q_i\rangle\}$ to calculate the transition
probabilities in Eqs.~(\ref{eq1}), (\ref{eq2}) and (\ref{eq3}).

Substituting the orthonormal basis of Eq.~(\ref{eq7}) into
Eqs~(\ref{eq2}), we get $\pi_{1,1}(t)$ and $\pi_{2,1}(t)$ as
follows,
\begin{equation}\label{eq8}
\pi_{1,1}(t)=\frac{N^2-2N+2}{N^2}+\frac{2(N-1)}{N^2}\cos Nt.
\end{equation}
\begin{equation}\label{eq9}
\pi_{2,1}(t)=\frac{2}{N^2}-\frac{2}{N^2}\cos Nt.
\end{equation}
For $\pi_{2,2}(t)$ and $\pi_{3,2}(t)$, the calculation is analogous,
but the expressions are cumbersome:
\begin{equation}\label{eq10}
\begin{array}{ll}
\pi_{2,2}(t)&=\frac{1}{N^2(N-1)^2}[(N^4-4N^3+5N^2-2N+2) \\
&+(2N^3-6N^2+4N)\cos t \\
&+(2N^2-4N)\cos (N-1)t \\
&+ (2N-2)\cos Nt].
\end{array}
\end{equation}
\begin{equation}\label{eq11}
\begin{array}{ll}
\pi_{3,2}(t)&=\frac{2}{N^2(N-1)^2}[(N^2-N+1) \\
&+(N-N^2)\cos t-N\cos (N-1)t\\
&+(N-1)\cos Nt].
\end{array}
\end{equation}
Other transition probabilities can also calculated, but they have
the same expressions as Eqs~(8)$\sim$(11). For instance,
$\pi_{4,2}(t)$ has the same analytical form as $\pi_{3,2}$, which is
consistent with our intuition. Analogously, we can get the long time
averages of the transition probabilities:
\begin{equation}\label{eq12}
\begin{array}{ll}
\chi_{1,1}=(N^2-2N+2)/N^2\\
\chi_{2,1}=2/N^2\\
\chi_{2,2}=(N^4-4N^3+5N^2-2N+2)/N^2/(N-1)^2\\
\chi_{3,2}=2(N^2-N+1)/N^2/(N-1)^2.
\end{array}
\end{equation}

The transition probabilities depend on the size of the graph. In the
thermodynamic limit of infinite network $N\rightarrow \infty$, the
transport displays high localizations on the initial position, i.e.,
$\chi_{i,j}\approx \delta_{i,j}$. On the contrary, for the classical
transport modeled by continuous-time random walks, the transition
probabilities do not show any oscillation and approach to the
equal-partition $1/N$ at long times~\cite{add-ref}. According to
Eq.~(\ref{eq1}), the four type transition probabilities can be
written as,
\begin{equation}\label{eq13}
\begin{array}{ll}
p_{1,1}(t)=1/N+e^{-Nt}(N-1)/N \\
p_{2,1}(t)=1/N-e^{-Nt}/N\\
p_{2,2}(t)=1/N+(N-2)/(N-1)e^{-t}+e^{-Nt}/N/(N-1)\\
p_{3,2}(t)=1/N-e^{-t}/(N-1)+e^{-Nt}/N/(N-1).
\end{array}
\end{equation}
\begin{figure}
\scalebox{0.75}[0.75]{\includegraphics{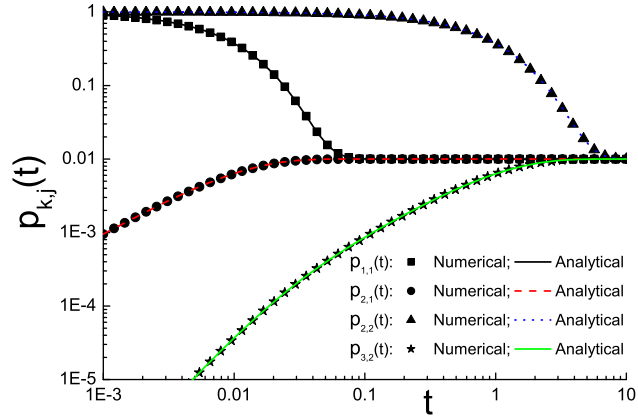}} \caption{(Color
online)Classical transition probabilities $p_{k,j}(t)$ versus $t$.
The marked points are numerical results and the curves are
analytical predictions in Eq.~(\ref{eq13}). \label{fg1}}
\end{figure}

In order to test the analytical predictions, we compare the
classical $p_{k,j}(t)$ predicted by Eq.~(\ref{eq13}) with the
numerical results obtained by numerically diagonalizing the
Hamiltonian. The results for a star graph of $N=100$ are shown in
Fig.~\ref{fg1}. As we can see, the numerical results exactly agree
with the analytical prediction in Eq.~(\ref{eq13}). The transport
reaches the equal-partitioned distribution $1/N$ at long times.
However, when the excitation starts at the central node, the
transport reaches the equal-partition more quickly than the
transport which starts at the leaf nodes (Compare the curves in
Fig.~\ref{fg1}).
\begin{figure}
\scalebox{0.8}[0.8]{\includegraphics{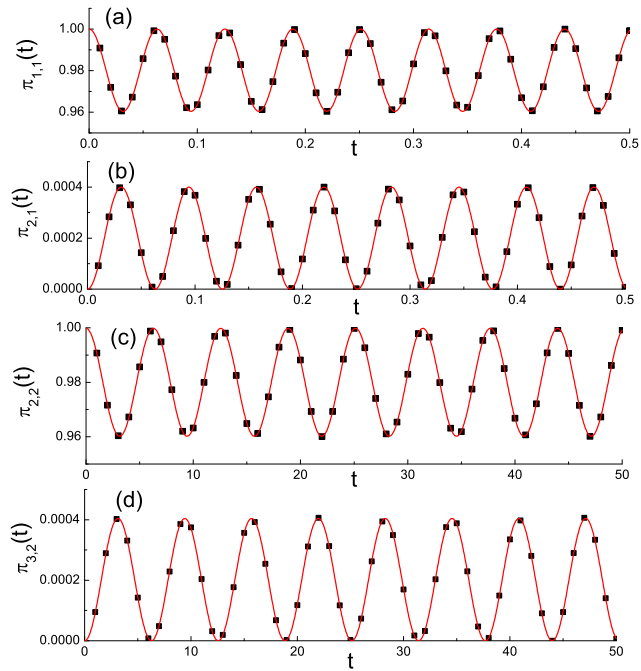}} \caption{(Color
online)Quantum transition probabilities $\pi_{1,1}(t)$ (a),
$\pi_{2,1}(t)$ (b), $\pi_{2,2}(t)$ (c) and $\pi_{3,2}(t)$ (d). The
points denoted by black squares are numerical results and the curves
are theoretical predictions in Eqs.~(\ref{eq8})-(\ref{eq11}).
\label{fg2}}
\end{figure}

For the quantum transport, we compare the transition probabilities
in Fig.~\ref{fg2}. The numerical results (marked as black squares)
exactly agree with the theoretical results in
Eqs.~(\ref{eq8})-(\ref{eq11}). We note that all the transition
probabilities show periodic recurrences. Comparing $\pi_{1,1}(t)$
and $\pi_{2,2}(t)$ (See Fig.~\ref{fg2} (a) and (c)), we find that
there are high probabilities to find the exciton at the initial
node. This suggests that the coherent transport shows high
localizations on the initial nodes~\cite{add-ref}. The oscillation
amplitudes of the return probabilities $\pi_{1,1}(t)$ and
$\pi_{2,2}(t)$ are comparable but the oscillation periods are quite
different. The oscillating period of $\pi_{2,2}(t)$ is $100$ (N)
times of that of $\pi_{1,1}(t)$. This could be interpreted by the
analytical expressions in Eqs.~(\ref{eq8}) and (\ref{eq10}). Similar
behavior also holds for $\pi_{2,1}(t)$ and $\pi_{3,2}(t)$ (See
Fig.~\ref{fg2} (b) and (d)), but the oscillation amplitude is
smaller than the return probabilities. This also can be understood
from the analytical results in Eqs.~(\ref{eq9}) and (\ref{eq11}),
where the transition probability is mainly determined by the high
order term of $N$. The small value of oscillating period of
$\pi_{2,1}(t)$ and $\pi_{1,1}(t)$ suggests that there are frequent
revivals when the exciton starts at the central node, compared to
the transport starting at the leaf nodes.
\begin{figure}
\scalebox{0.7}[0.7]{\includegraphics{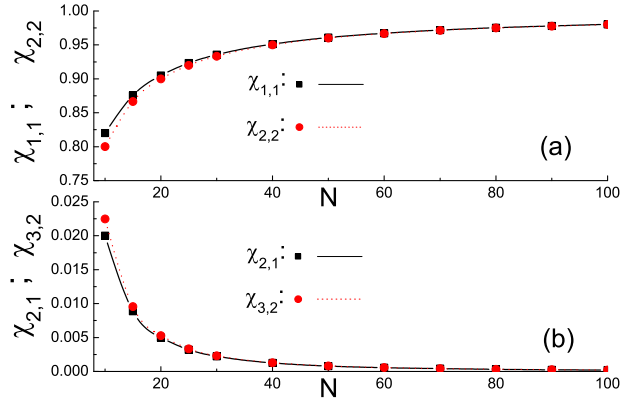}} \caption{(Color
online)Long-time limiting probabilities $\chi_{1,1}$, $\chi_{2,2}$
(a) and $\chi_{2,1}$, $\chi_{3,2}$ (b) as a function of the network
size $N$. The points marked as symbols are numerical results and the
curves are analytical results predicted by
Eq.~(\ref{eq12}).\label{fg3}}
\end{figure}

The quantum limiting probabilities in Eq.~(\ref{eq12}) are only a
function of graph size $N$. Fig.~\ref{fg3} shows the quantum
limiting probabilities for numerical results and theoretical
predictions. Both the results agree with each other. We find that
the return probabilities $\chi_{1,1}$ and $\chi_{2,2}$ are an
incremental function of $N$ and approach to $1$ in the limit $N\to
\infty$. By contraries, $\chi_{2,1}$ and $\chi_{3,2}$ decrease with
$N$ and close to $2/N^2$ in the limit $N\to \infty$. We note that
$\chi_{1,1}$ differs from $\chi_{2,2}$ for small values of $N$. Such
deviation diminishes as $N$ increases. This suggests that the
strength of localizations is almost the same for central-node and
leaf-node excitations. The only difference is that the frequency of
revivals (oscillation period) for central node excitation is much
higher than that for leaf node excitation.

To address the similarity and difference between the star graph and
complete graph, we proceed to consider the transport on a complete
graph of size $N$. The complete graph is fully connected, thus the
Hamiltonian is given by $H=(N-1)\sum_i|i\rangle \langle
i|-\sum_{i\neq j}|i\rangle \langle j|$. The eigenvalues are two
different values: $E_1=E_2=...=E_{N-1}=N$ and $E_N=0$. One set of
un-orthogonal states $\{|v_i\rangle\}$ corresponding to the
eigenvalues can be written as: $|v_i\rangle =|i+1\rangle -|1\rangle$
($i=1,2,...,N-1$) and $|v_{N}\rangle=\sum_{j=1}^N |j\rangle$. Using
the Gram-Schmidt orthonormalization (See Eq.~(\ref{eq5})), the
orthonormal basis for a complete graph is,
\begin{equation}\label{eq14}
|q_i\rangle=\left\{
\begin{array}{lll}
\sqrt{\frac{i}{i+1}}|i+1\rangle-\sqrt{\frac{1}{i(i+1)}}\sum_{j=1}^{i}|j\rangle,   &  i=1,2,...,N-1 \\
\sqrt{1/N}\sum_{j=1}^N |j\rangle,  & i=N.\\
\end{array}
\right.
\end{equation}
Substituting the above Equation into Eq.~(\ref{eq2}), we get the
quantum transition probabilities for complete graph,
\begin{equation}\label{eq15}
\pi_{i,j}(t)=\left\{
\begin{array}{lll}
\frac{N^2-2N+2}{N^2}+\frac{2(N-1)}{N^2}\cos Nt,   &  i=j \\
\frac{2}{N^2}-\frac{2}{N^2}\cos Nt.  & i\neq j.\\
\end{array}
\right.
\end{equation}

We note that Eq.~(\ref{eq15}) is exactly the same form as
Eqs.~(\ref{eq8}) and (\ref{eq9}). This indicates that the transport
starting at central node on star graph is equivalent to the
transport on a complete graph of the same size.

In summary, we have studied coherent exciton transport of
continuous-time quantum walks on star graph. Exact analytical
results of the transition probabilities are obtained in terms of the
Gram-Schmidt orthonormalization. We find that the coherent transport
shows perfect recurrences and there are high frequency of revivals
for central node excitation. Study of long time averages suggests
that the quantum transport displays strong localizations on the
initial node. When the initial excitation starts at the central
node, the transport on star graph is equivalent to the transport on
a complete graph of the same size.

This work is supported by National Natural Science Foundation of
China under projects 10575042, 10775058 and MOE of China under
contract number IRT0624 (CCNU).

\end{document}